\expandafter \def \csname CHAPLABELintro\endcsname {1}
\expandafter \def \csname CHAPLABELconjecture\endcsname {2}
\expandafter \def \csname TABLABELrank\endcsname {2.1?}
\expandafter \def \csname TABLABELrank2\endcsname {2.2?}
\expandafter \def \csname CHAPLABELeg\endcsname {3}
\expandafter \def \csname CHAPLABELfin\endcsname {4}

\font\eightrm=cmr8 at 8pt

\font\seventeenrm=cmr17 at 17pt
\font\twentyonerm=cmr17 at 21pt

\font\ss=cmss10

\font\csc=cmcsc10

\font\twelvecal=cmsy10 at 12pt

\font\twelvemath=cmmi12

\font\seventeenbold=cmbx7 at 17pt

\font\fively=lasy5
\font\sevenly=lasy7
\font\tenly=lasy10

\textfont10=\tenly
\scriptfont10=\sevenly
\scriptscriptfont10=\fively
\magnification=1200
\parskip=10pt
\parindent=20pt
\def\today{\ifcase\month\or January\or February\or March\or April\or May\or
June
       \or July\or August\or September\or October\or November\or December\fi
       \space\number\day, \number\year}

\def\title#1{\footline={\ifnum\pageno<2\hfil
       \else\hss\tenrm\folio\hss\fi}\vskip1truein\centerline{{#1}
       \footnote{\raise1ex\hbox{*}}{\eightrm Supported in part
       by the Robert A. Welch Foundation and N.S.F. Grants
       PHY-880637 and\break PHY-8605978.}}}

\def\newpage{\vfill\eject}
\def\abstract#1{\centerline{\bf ABSTRACT}\vskip.2truein{\narrower\noindent#1
       \smallskip}}

\def\runninghead#1#2{\voffset=2\baselineskip\nopagenumbers
       \headline={\ifodd\pageno\rightheadline\else \leftheadline\fi}
       \def\rightheadline{{\sl#1}\hfill{\rm\folio}}
       \def\leftheadline{{\rm\folio}\hfill{\sl#2}}}
\def\SS{\mathhexbox278}

\newcount\footnoteno
\def\Footnote#1{\advance\footnoteno by 1
                \let\SF=\empty
                \ifhmode\edef\SF{\spacefactor=\the\spacefactor}\/\fi
                $^{\the\footnoteno}$\ignorespaces
                \SF\vfootnote{$^{\the\footnoteno}$}{#1}}

\def\place#1#2#3{\vbox to0pt{\kern-\parskip\kern-7pt
                             \kern-#2truein\hbox{\kern#1truein #3}
                             \vss}\nointerlineskip}
\def\figurecaption#1#2{\kern.75truein\vbox{\hsize=5truein\noindent{\bf Figure
    \figlabel{#1}:} #2}}
\def\tablecaption#1#2{\kern.75truein\lower12truept\hbox{\vbox{\hsize=5truein
    \noindent{\bf Table\hskip5truept\tablabel{#1}:} #2}}}
\def\boxed#1{\lower3pt\hbox{
                       \vbox{\hrule\hbox{\vrule

\vbox{\kern2pt\hbox{\kern3pt#1\kern3pt}\kern3pt}\vrule}
                         \hrule}}}

\def\g{\gamma}
\def\D{\Delta}

\def\l{\lambda}

\def\S{\Sigma}
\def\t{\tau}

\def\ca#1{\relax\ifmmode {{\cal #1}}\else $\cal #1$\fi}

\def\calb{{\cal B}}

\def\calm{{\cal M}}

\def\inbar{\vrule height1.5ex width.4pt depth0pt}
\def\IB{\relax{\rm I\kern-.18em B}}
\def\IC{\relax\hbox{\kern.25em$\inbar\kern-.3em{\rm C}$}}
\def\ID{\relax{\rm I\kern-.18em D}}
\def\IE{\relax{\rm I\kern-.18em E}}
\def\IF{\relax{\rm I\kern-.18em F}}
\def\IG{\relax\hbox{\kern.25em$\inbar\kern-.3em{\rm G}$}}
\def\IH{\relax{\rm I\kern-.18em H}}
\def\II{\relax{\rm I\kern-.18em I}}
\def\IK{\relax{\rm I\kern-.18em K}}
\def\IL{\relax{\rm I\kern-.18em L}}
\def\IM{\relax{\rm I\kern-.18em M}}
\def\IN{\relax{\rm I\kern-.18em N}}
\def\IO{\relax\hbox{\kern.25em$\inbar\kern-.3em{\rm O}$}}
\def\IP{\relax{\rm I\kern-.18em P}}
\def\IQ{\relax\hbox{\kern.25em$\inbar\kern-.3em{\rm Q}$}}
\def\IR{\relax{\rm I\kern-.18em R}}
\def\IZ{\relax\ifmmode\hbox{\ss Z\kern-.4em Z}\else{\ss Z\kern-.4em Z}\fi}
\def\IGa{\relax{\rm I}\kern-.18em\Gamma}
\def\IPi{\relax{\rm I}\kern-.18em\Pi}
\def\ITh{\relax\hbox{\kern.25em$\inbar\kern-.3em\Theta$}}
\def\IOm{\relax\thinspace\inbar\kern1.95pt\inbar\kern-5.525pt\Omega}


\def\ie{{\it i.e.,\ \/}}
\def\eg{{\it e.g.,\ \/}}

\def\cy{Calabi--Yau}
\def\cym{Calabi--Yau manifold}

\def\cyt{Calabi--Yau threefold}

\def\H#1#2{\relax\ifmmode {H^{#1#2}}\else $H^{#1 #2}$\fi}
\def\M{\relax\ifmmode{\calm}\else $\calm$\fi}

\def\Bigcheck{\lower3.8pt\hbox{\smash{\hbox{{\twentyonerm \v{}}}}}}
\def\bigboldcheck{\smash{\hbox{{\seventeenbold\v{}}}}}

\def\Bighat{\lower3.8pt\hbox{\smash{\hbox{{\twentyonerm \^{}}}}}}

\def\Msharp{\relax\ifmmode{\calm^\sharp}\else $\smash{\calm^\sharp}$\fi}
\def\Mflat{\relax\ifmmode{\calm^\flat}\else $\smash{\calm^\flat}$\fi}
\def\preMcheck{\kern2pt\hbox{\Bigcheck\kern-12pt{$\cal M$}}}
\def\Mcheck{\relax\ifmmode\preMcheck\else $\preMcheck$\fi}
\def\preMhat{\kern2pt\hbox{\Bighat\kern-12pt{$\cal M$}}}
\def\Mhat{\relax\ifmmode\preMhat\else $\preMhat$\fi}

\def\Bsharp{\relax\ifmmode{\calb^\sharp}\else $\calb^\sharp$\fi}
\def\Bflat{\relax\ifmmode{\calb^\flat}\else $\calb^\flat$ \fi}
\def\preBcheck{\hbox{\Bigcheck\kern-9pt{$\cal B$}}}
\def\Bcheck{\relax\ifmmode\preBcheck\else $\preBcheck$\fi}
\def\preBhat{\hbox{\Bighat\kern-9pt{$\cal B$}}}
\def\Bhat{\relax\ifmmode\preBhat\else $\preBhat$\fi}

\def\figBcheck{\kern3pt\hbox{\raise1pt\hbox{\bigboldcheck}\kern-11pt
    {\twelvecal B}}}
\def\figBsharp{{\twelvecal B}\raise5pt\hbox{$\twelvemath\sharp$}}
\def\figBflat{{\twelvecal B}\raise5pt\hbox{$\twelvemath\flat$}}

\def\gcheck{\hbox{\lower2.5pt\hbox{\Bigcheck}\kern-8pt$\g$}}
\def\lhat{\hbox{\raise.5pt\hbox{\Bighat}\kern-8pt$\l$}}

\def\Fcheck{\kern2pt\hbox{\raise1pt\hbox{\Bigcheck}\kern-10pt{$\cal F$}}}
\def\Fhat{\kern2pt\hbox{\raise1pt\hbox{\Bighat}\kern-10pt{$\cal F$}}}

\def\cp#1{\relax\ifmmode {\IP\kern-2pt{}_{#1}}\else $\IP\kern-2pt{}_{#1}$\fi}
\def\h#1#2{\relax\ifmmode {b_{#1#2}}\else $b_{#1#2}$\fi}

\def\frac#1#2{{#1\over #2}}

\def\cone{\relax\thinspace\hbox{$<\kern-.8em{)}$}}
\mathchardef\mho"0A30

\def\-{\hphantom{-}}


\def\npb#1{Nucl.\ Phys.\ {\bf B#1}}

\def\plb#1{Phys. Lett. {\bf #1B}}


\def\picture #1 by #2 (#3){\vbox to #2{\hrule width #1 height 0pt depth 0pt
                                       \vfill\special{picture #3}}}
\def\scaledpicture #1 by #2 (#3 scaled #4){{\dimen0=#1 \dimen1=#2
           \divide\dimen0 by 1000 \multiply\dimen0 by #4
            \divide\dimen1 by 1000 \multiply\dimen1 by #4
            \picture \dimen0 by \dimen1 (#3 scaled #4)}}
\def\illustration #1 by #2 (#3){\vbox to #2{\hrule width #1 height 0pt depth
0pt
                                       \vfill\special{illustration #3}}}
\def\scaledillustration #1 by #2 (#3 scaled #4){{\dimen0=#1 \dimen1=#2
           \divide\dimen0 by 1000 \multiply\dimen0 by #4
            \divide\dimen1 by 1000 \multiply\dimen1 by #4
            \illustration \dimen0 by \dimen1 (#3 scaled #4)}}


\def\delaOssa{\nobreak\vskip1truein\hbox to\hsize
       {\hskip 4truein Xenia de la Ossa\hfill}}

\def\hoy{\number\day\space de \ifcase\month\or enero\or febrero\or marzo\or
       abril\or mayo\or junio\or julio\or agosto\or septiembre\or octubre\or
       noviembre\or diciembre\fi\space de \number\year}


\newif\ifproofmode
\proofmodefalse

\newif\ifforwardreference
\forwardreferencefalse

\newif\ifchapternumbers
\chapternumbersfalse

\newif\ifcontinuousnumbering
\continuousnumberingfalse

\newif\iffigurechapternumbers
\figurechapternumbersfalse

\newif\ifcontinuousfigurenumbering
\continuousfigurenumberingfalse

\newif\iftablechapternumbers
\tablechapternumbersfalse

\newif\ifcontinuoustablenumbering
\continuoustablenumberingfalse

\font\eqsixrm=cmr6

\def\marginstyle{\eqsixrm}

\newtoks\chapletter
\newcount\chapno
\newcount\eqlabelno
\newcount\figureno
\newcount\tableno

\chapno=0
\eqlabelno=0
\figureno=0
\tableno=0

\def\chapfolio{\ifnum\chapno>0 \the\chapno\else\the\chapletter\fi}

\def\bumpchapno{\ifnum\chapno>-1 \global\advance\chapno by 1
\else\global\advance\chapno by -1 \setletter\chapno\fi
\ifcontinuousnumbering\else\global\eqlabelno=0 \fi
\ifcontinuousfigurenumbering\else\global\figureno=0 \fi
\ifcontinuoustablenumbering\else\global\tableno=0 \fi}

\def\setletter#1{\ifcase-#1{}\or{}%
\or\global\chapletter={A}%
\or\global\chapletter={B}%
\or\global\chapletter={C}%
\or\global\chapletter={D}%
\or\global\chapletter={E}%
\or\global\chapletter={F}%
\or\global\chapletter={G}%
\or\global\chapletter={H}%
\or\global\chapletter={I}%
\or\global\chapletter={J}%
\or\global\chapletter={K}%
\or\global\chapletter={L}%
\or\global\chapletter={M}%
\or\global\chapletter={N}%
\or\global\chapletter={O}%
\or\global\chapletter={P}%
\or\global\chapletter={Q}%
\or\global\chapletter={R}%
\or\global\chapletter={S}%
\or\global\chapletter={T}%
\or\global\chapletter={U}%
\or\global\chapletter={V}%
\or\global\chapletter={W}%
\or\global\chapletter={X}%
\or\global\chapletter={Y}%
\or\global\chapletter={Z}\fi}

\def\tempsetletter#1{\ifcase-#1{}\or{}%
\or\global\chapletter={A}%
\or\global\chapletter={B}%
\or\global\chapletter={C}%
\or\global\chapletter={D}%
\or\global\chapletter={E}%
\or\global\chapletter={F}%
\or\global\chapletter={G}%
\or\global\chapletter={H}%
\or\global\chapletter={I}%
\or\global\chapletter={J}%
\or\global\chapletter={K}%
\or\global\chapletter={L}%
\or\global\chapletter={M}%
\or\global\chapletter={N}%
\or\global\chapletter={O}%
\or\global\chapletter={P}%
\or\global\chapletter={Q}%
\or\global\chapletter={R}%
\or\global\chapletter={S}%
\or\global\chapletter={T}%
\or\global\chapletter={U}%
\or\global\chapletter={V}%
\or\global\chapletter={W}%
\or\global\chapletter={X}%
\or\global\chapletter={Y}%
\or\global\chapletter={Z}\fi}

\def\chapshow#1{\ifnum#1>0 \relax#1%
\else{\tempsetletter{\number#1}\chapno=#1\chapfolio}\fi}

\def\chaplabel#1{\bumpchapno\ifproofmode\ifforwardreference
\immediate\write1{\noexpand\expandafter\noexpand\def
\noexpand\csname CHAPLABEL#1\endcsname{\the\chapno}}\fi\fi
\global\expandafter\edef\csname CHAPLABEL#1\endcsname
{\the\chapno}\ifproofmode\llap{\hbox{\marginstyle #1\ }}\fi\chapfolio}

\def\eqnum{\global\advance\eqlabelno by 1
\eqno(\ifchapternumbers\chapfolio.\fi\the\eqlabelno)}

\def\eqlabel#1{\global\advance\eqlabelno by 1 \ifproofmode\ifforwardreference
\immediate\write1{\noexpand\expandafter\noexpand\def
\noexpand\csname EQLABEL#1\endcsname{\the\chapno.\the\eqlabelno?}}\fi\fi
\global\expandafter\edef\csname EQLABEL#1\endcsname
{\the\chapno.\the\eqlabelno?}\eqno(\ifchapternumbers\chapfolio.\fi
\the\eqlabelno)\ifproofmode\rlap{\hbox{\marginstyle #1}}\fi}

\def\eqalignnum{\global\advance\eqlabelno by 1
&(\ifchapternumbers\chapfolio.\fi\the\eqlabelno)}

\def\eqalignlabel#1{\global\advance\eqlabelno by 1 \ifproofmode
\ifforwardreference\immediate\write1{\noexpand\expandafter\noexpand\def
\noexpand\csname EQLABEL#1\endcsname{\the\chapno.\the\eqlabelno?}}\fi\fi
\global\expandafter\edef\csname EQLABEL#1\endcsname
{\the\chapno.\the\eqlabelno?}&(\ifchapternumbers\chapfolio.\fi
\the\eqlabelno)\ifproofmode\rlap{\hbox{\marginstyle #1}}\fi}

\def\eqref#1{\hbox{(\ifundefined{EQLABEL#1}***)\ifproofmode\ifforwardreference%
\else\write16{ ***Undefined Equation Reference #1*** }\fi
\else\write16{ ***Undefined Equation Reference #1*** }\fi
\else\edef\LABxx{\getlabel{EQLABEL#1}}%
\def\LAByy{\expandafter\stripchap\LABxx}\ifchapternumbers%
\chapshow{\LAByy}.\expandafter\stripeq\LABxx%
\else\ifnum\number\LAByy=\chapno\relax\expandafter\stripeq\LABxx%
\else\chapshow{\LAByy}.\expandafter\stripeq\LABxx\fi\fi)\fi}%
\ifproofmode\write2{Equation #1}\fi}

\def\fignum{\global\advance\figureno by 1
\relax\iffigurechapternumbers\chapfolio.\fi\the\figureno}

\def\figlabel#1{\global\advance\figureno by 1
\relax\ifproofmode\ifforwardreference
\immediate\write1{\noexpand\expandafter\noexpand\def
\noexpand\csname FIGLABEL#1\endcsname{\the\chapno.\the\figureno?}}\fi\fi
\global\expandafter\edef\csname FIGLABEL#1\endcsname
{\the\chapno.\the\figureno?}\iffigurechapternumbers\chapfolio.\fi
\ifproofmode\llap{\hbox{\marginstyle#1
\kern1.2truein}}\relax\fi\the\figureno}

\def\figref#1{\hbox{\ifundefined{FIGLABEL#1}!!!!\ifproofmode\ifforwardreference%
\else\write16{ ***Undefined Figure Reference #1*** }\fi
\else\write16{ ***Undefined Figure Reference #1*** }\fi
\else\edef\LABxx{\getlabel{FIGLABEL#1}}%
\def\LAByy{\expandafter\stripchap\LABxx}\iffigurechapternumbers%
\chapshow{\LAByy}.\expandafter\stripeq\LABxx%
\else\ifnum \number\LAByy=\chapno\relax\expandafter\stripeq\LABxx%
\else\chapshow{\LAByy}.\expandafter\stripeq\LABxx\fi\fi\fi}%
\ifproofmode\write2{Figure #1}\fi}

\def\tabnum{\global\advance\tableno by 1
\relax\iftablechapternumbers\chapfolio.\fi\the\tableno}

\def\tablabel#1{\global\advance\tableno by 1
\relax\ifproofmode\ifforwardreference
\immediate\write1{\noexpand\expandafter\noexpand\def
\noexpand\csname TABLABEL#1\endcsname{\the\chapno.\the\tableno?}}\fi\fi
\global\expandafter\edef\csname TABLABEL#1\endcsname
{\the\chapno.\the\tableno?}\iftablechapternumbers\chapfolio.\fi
\ifproofmode\llap{\hbox{\marginstyle#1
\kern1.2truein}}\relax\fi\the\tableno}

\def\tabref#1{\hbox{\ifundefined{TABLABEL#1}!!!!\ifproofmode\ifforwardreference%
\else\write16{ ***Undefined Table Reference #1*** }\fi
\else\write16{ ***Undefined Table Reference #1*** }\fi
\else\edef\LABtt{\getlabel{TABLABEL#1}}%
\def\LABTT{\expandafter\stripchap\LABtt}\iftablechapternumbers%
\chapshow{\LABTT}.\expandafter\stripeq\LABtt%
\else\ifnum\number\LABTT=\chapno\relax\expandafter\stripeq\LABtt%
\else\chapshow{\LABTT}.\expandafter\stripeq\LABtt\fi\fi\fi}%
\ifproofmode\write2{Table#1}\fi}

\newdimen\sectionskip     \sectionskip=20truept
\newcount\sectno
\def\section#1#2{\sectno=0 \null\vskip\sectionskip
    \centerline{\chaplabel{#1}.~~{\bf#2}}\nobreak\vskip.2truein
    \noindent\ignorespaces}

\def\advancesectno{\global\advance\sectno by 1}
\def\sectfolio{\number\sectno}
\def\subsection#1{\goodbreak\advancesectno\null\vskip10pt
                  \noindent\chapfolio.~\sectfolio.~{\bf #1}
                  \nobreak\vskip.05truein\noindent\ignorespaces}

\def\uttg#1{\null\vskip.1truein
    \ifproofmode \line{\hfill{\bf Draft}:
    UTTG--{#1}--\number\year}\line{\hfill\today}
    \else \line{\hfill UTTG--{#1}--\number\year}
    \line{\hfill\ifcase\month\or January\or February\or March\or April\or
May\or June
    \or July\or August\or September\or October\or November\or December\fi
    \space\number\year}\fi}

\def\getlabel#1{\csname#1\endcsname}
\def\ifundefined#1{\expandafter\ifx\csname#1\endcsname\relax}
\def\stripchap#1.#2?{#1}
\def\stripeq#1.#2?{#2}

%
\catcode`@=11 
\def\space@ver#1{\let\@sf=\empty\ifmmode#1\else\ifhmode%
\edef\@sf{\spacefactor=\the\spacefactor}\unskip${}#1$\relax\fi\fi}
\newcount\referencecount     \referencecount=0
\newif\ifreferenceopen       \newwrite\referencewrite
\newtoks\rw@toks
\def\refmark#1{\relax[#1]}
\def\refend{\refmark{\number\referencecount}}
\newcount\lastrefsbegincount \lastrefsbegincount=0
\def\refsend{\refmark{\count255=\referencecount%
\advance\count255 by -\lastrefsbegincount%
\ifcase\count255 \number\referencecount%
\or\number\lastrefsbegincount,\number\referencecount%
\else\number\lastrefsbegincount-\number\referencecount\fi}}
\def\refch@ck{\chardef\rw@write=\referencewrite
\ifreferenceopen\else\referenceopentrue
\immediate\openout\referencewrite=referenc.texauxil \fi}
%
{\catcode`\^^M=\active 
  \gdef\obeyendofline{\catcode`\^^M\active \let^^M\ }}%
%
{\catcode`\^^M=\active 
  \gdef\ignoreendofline{\catcode`\^^M=5}}
{\obeyendofline\gdef\rw@start#1{\def\t@st{#1}\ifx\t@st\blankend%
\endgroup\@sf\relax\else\ifx\t@st\bl@nkend\endgroup\@sf\relax%
\else\rw@begin#1
\backtotext
\fi\fi}}
{\obeyendofline\gdef\rw@begin#1
{\def\n@xt{#1}\rw@toks={#1}\relax%
\rw@next}}
\def\blankend{}
{\obeylines\gdef\bl@nkend{
}}
\newif\iffirstrefline  \firstreflinetrue
\def\rwr@teswitch{\ifx\n@xt\blankend\let\n@xt=\rw@begin%
\else\iffirstrefline\global\firstreflinefalse%
\immediate\write\rw@write{\noexpand\obeyendofline\the\rw@toks}%
\let\n@xt=\rw@begin%
\else\ifx\n@xt\rw@@d \def\n@xt{\immediate\write\rw@write{%
\noexpand\ignoreendofline}\endgroup\@sf}%
\else\immediate\write\rw@write{\the\rw@toks}%
\let\n@xt=\rw@begin\fi\fi\fi}
\def\rw@next{\rwr@teswitch\n@xt}
\def\rw@@d{\backtotext} \let\rw@end=\relax
\let\backtotext=\relax

\newdimen\refindent     \refindent=30pt
\def\Textindent#1{\noindent\llap{#1\enspace}\ignorespaces}
\def\refitem#1{\par\hangafter=0 \hangindent=\refindent\Textindent{#1}}
\def\REFNUM#1{\space@ver{}\refch@ck\firstreflinetrue%
\global\advance\referencecount by 1 \xdef#1{\the\referencecount}}
\def\refnum#1{\space@ver{}\refch@ck\firstreflinetrue%
\global\advance\referencecount by 1\xdef#1{\the\referencecount}\refend}

\def\REF#1{\REFNUM#1%
\immediate\write\referencewrite{%
\noexpand\refitem{#1.}}%
\begingroup\obeyendofline\rw@start}
\def\ref{\refnum\?%
\immediate\write\referencewrite{\noexpand\refitem{\?.}}%
\begingroup\obeyendofline\rw@start}
\def\Ref#1{\refnum#1%
\immediate\write\referencewrite{\noexpand\refitem{#1.}}%
\begingroup\obeyendofline\rw@start}
\def\REFS#1{\REFNUM#1\global\lastrefsbegincount=\referencecount%
\immediate\write\referencewrite{\noexpand\refitem{#1.}}%
\begingroup\obeyendofline\rw@start}

\def\REFSCON#1{\REF#1}

\def\cite#1{\refmark#1}
\catcode`@=12 
%
\input epsf.tex
\baselineskip=15pt plus 1pt minus 1pt
\parskip=5pt
\chapternumberstrue
\figurechapternumberstrue
\tablechapternumberstrue
\ifproofmode
\immediate\openout2=allcrossreferfile \fi
\ifforwardreference\input labelfile
\ifproofmode\immediate\openout1=labelfile \fi\fi
\hfuzz=2pt
\vfuzz=2pt


\def\hourandminute{\count255=\time\divide\count255 by 60
\xdef\hour{\number\count255}
\multiply\count255 by -60\advance\count255 by\time
\hour:\ifnum\count255<10 0\fi\the\count255}

\def\subsection#1{\goodbreak\advancesectno\null\vskip10pt
                  \noindent{\it \chapfolio.\sectfolio.~#1}
                  \nobreak\vskip.05truein\noindent\ignorespaces}
\def\cite#1{\refmark{#1}}
\def\\{\hfill\break}

\def\point#1{\noindent\setbox0=\hbox{#1}\kern-\wd0\box0}

\def\etal{{\it et al.\/}}
\nopagenumbers\pageno=0
\rightline{\eightrm UTTG-19-97}\vskip-5pt
\rightline{\eightrm hep-th/9706005}\vskip-5pt
\rightline{\eightrm June 1, 1997}

\vskip1.2truein
\centerline{\seventeenrm Mirror Symmetry via Deformation of Bundles on K3
Surfaces}
\vskip60pt
\centerline{\csc Eugene~Perevalov${^\ast}$ and 
Govindan~Rajesh$^{\dag}$}
\vfootnote{$^{\eightrm \ast}$}{\eightrm pereval@physics.utexas.edu}
\vfootnote{$^{\eightrm \dag}$}{\eightrm rajesh@physics.utexas.edu}

\vskip1truein\bigskip
\centerline{\it Theory Group}
\centerline{\it Department of Physics}
\centerline{\it University of Texas}
\centerline{\it Austin, TX 78712, USA}
\vskip.8in\bigskip
\nobreak\vbox{
\centerline{\bf ABSTRACT}
\vskip.25truein
\noindent{We consider F-theory compactifications on a mirror
pair of elliptic \cyt s. This yields two different six-dimensional theories,
each of them being nonperturbatively equivalent to some compactification of
heterotic strings on a $K3$ surface $S$ with certain bundle data 
$E\rightarrow S$. We find evidence for a transformation of $S$
together with the bundle that takes one heterotic model to the other. }
} 
\newpage
\pageno=1
\headline={\ifproofmode\hfil\eightrm draft:\ \today\
\hourandminute\else\hfil\fi}
\footline={\rm\hfil\folio\hfil}
\section{intro}{Introduction}
An area that has the been focus of extensive research in recent times is that
of Heterotic/Type II duality, which was first studied in~
\REFS\HT{C.~Hull and P.~K.~Townsend, \npb{438} (1995) 109, hep-th/9410167.}
\REFSCON\WitI{E.~Witten, \npb{443} (1995) 85, hep-th/9503124.}
\REFSCON\HS{J. A. Harvey and A. Strominger, \npb{449} (1995) 535, 
hep-th/9504047.}
\REFSCON\S{A. Sen, \npb{450} (1995) 103, hep-th/9504027.}
\refsend , where compactifications with 16 supersymmetry generators were 
considered. These results were extended to $N=2$, $d=4$ compactifications in~
\REFS\KV{S.~Kachru and C.~Vafa, \npb{450} (1995) (69), hep-th/9505105.}
\REFSCON\FNSV{S. Ferrara, J. A. Harvey, A. Strominger and C. Vafa,
\plb{361} (1995) 59,\\ hep-th/9505162.}
\refsend
, where the heterotic theory on $K3\times T^2$ was conjectured to be dual to
the Type IIA theory on a \cyt . This duality may be lifted to six dimensions
upon going to the large radius limit of the torus (with the Wilson lines, if
any, switched off), to obtain Heterotic/F-Theory duality in six dimensions~
\REFS\rMVI{D.~R.~Morrison and C.~Vafa, Nucl. Phys. {\bf B473} (1996) 74,
hep-th/9602114.}
\REFSCON\rMVII{D.~R.~Morrison and C.~Vafa, Nucl. Phys. {\bf B476} (1996) 437, 
hep-th/9603161.}
\refsend , where the \cyt\ is now an elliptic fibration. This raises an
interesting question. Consider F-theory compactifications on a mirror
pair of elliptic \cyt s. This yields two different six-dimensional theories,
each of them being nonperturbatively equivalent to some compactification of
heterotic strings on a $K3$ surface $S$ with certain bundle data 
$E\rightarrow S$. Is there a simple transformation of $S$ together with
the bundle that would take one heterotic model to the other?

This is the issue that we address in this note. We argue that given a
heterotic vacuum dual to F-theory compactified on an
elliptic \cyt\ \ca{M}, the heterotic dual of F-theory compactified on the
mirror manifold \ca{W} is obtained by essentially exchanging the roles of
large (\ie finite-sized) and small (\ie point-like) instantons. The
non-perturbative phenomena associated with the appearance of small instantons
were first studied in~
\Ref\WitII{E.~Witten, \npb{460} (1996) 541, hep-th/9511030.}\ and their results
have been extended in~ 
\REFS\SW{N. Seiberg and E. Witten, \npb{471} (1996) 121, hep-th/9603003.} 
\REFSCON\B{M. Berkooz \etal, \npb{475} (1996) 115, hep-th/9605184.}
\REFSCON\Asp{P. S. Aspinwall, hep-th/9612108.}
\REFSCON\Ken{K. Intriligator, hep-th/9702038.}
\REFSCON\KenBlumI{J. D. Blum and K. Intriligator, hep-th/9705030.}
\REFSCON\KenBlumII{J. D. Blum and K. Intriligator, hep-th/9705044.}
\refsend . In this paper, we will mainly be using the results of~
\Ref\AM{P.~S.~Aspinwall and D.~R.~Morrison, hep-th/9705104.}\ which describes
the enhanced gauge symmetry that results when small instantons coalesce onto
orbifold singularities of $K3$.

The organization of this note is as follows. In \SS{2}, we state our
proposal relating pairs of heterotic theories to mirror pairs of \cyt s,
and prove that the Hodge numbers are consistent with our hypothesis. In
\SS{3}, we study some examples of mirror pairs and verify that the gauge
and tensor multiplet content of the F-theory compactifications
(which is obtained from the
singularity structure of the \cyt s~
\REFS\BG{P.~Candelas, E.~Perevalov and G.~Rajesh, hep-th/9704097.}
\REFSCON\AG{P.~S.~Aspinwall and M.~Gross, unpublished.}
\REFSCON\Mat{P.~Candelas, E.~Perevalov and G.~Rajesh, to appear.}
\refsend ) matches that obtained on the heterotic side~\cite{\AM}.
\SS{4} summarises our results.
\newpage
\section{conjecture}{Deformations of bundles over $K3$ surfaces}
In this section, we describe our proposal which relates F-theory
compactifications on mirror pairs of \cyt s and their heterotic duals. Our
results are valid for elliptic \cyt s whose mirrors are also elliptic
fibrations.
\subsection{The Proposal}
Heterotic string theory compactified on an elliptic $K3$, with some
appropriate choice of gauge bundle, yields a six dimensional theory with $N=1$
supersymmetry. This is conjectured to be dual to F-theory compactified on an
elliptic \cyt\ \ca{M}. Now consider F-theory compactified on the mirror
manifold \ca{W}. Our proposal is that the heterotic dual of this model can be
obtained from the original heterotic model by applying the following map:
\item{$\bullet$} Large instantons in a gauge bundle with structure group $H$
map to
an equal number of small instantons sitting on a {\ss H} orbifold singularity
of the $K3$ and vice versa, where {\ss H} denotes both the subgroup of
$SL(2,\IZ )$ as well as the surface singularity corresponding to $H$.
\item{$\bullet$} Small instantons sitting on a {\sl smooth\/} point of the $K3$
map to themselves.

\noindent For example, the heterotic $E_8\times E_8$ theory compactified on
a $K3$ with
24 (large) instantons in an $E_8$ gauge bundle yields a six-dimensional theory
with generic gauge group $E_8$. Its F-theory dual is obtained using the \cyt\
\ca{M} with Hodge numbers $h_{11}=11$, $h_{21}=491$. Compactifying F-theory on
the mirror \ca{W} of this manifold yields a theory with 193 tensor
multiplets and gauge group
$E_8^{17}{\times}F_4^{16}{\times}G_2^{32}{\times}SU(2)^{32}$~
\cite{\BG,\AG}. Using the map
described above, we find that the heterotic dual is compactified on a $K3$
with 24 small instantons on an {\ss E}$_8$ singularity, which was shown in
Ref.~\cite{\AM}.  
\subsection{Evidence for the proposal: matching Hodge numbers}
The necessary condition for our proposal to work is that the heterotic theory
obtained by applying the above map be dual to F-theory compactified on a
manifold \ca{W'} with Hodge numbers which are precisely those of \ca{W}, \ie
$h_{11}(\ca{W'})=h_{21}(\ca{M})$ and $h_{21}(\ca{W'})=h_{11}(\ca{M})$.  

This can be proved as follows. 
For definiteness, let us suppose that we start
with a heterotic $E_8\times E_8$ model in which all the instantons are large
and sitting in a bundle with structure group $H\times H'$ (where $H$ and $H'$
are subgroups of the first and second $E_8$ respectively), so that the
gauge symmetry is $G\times G'$, where $G$ and $G'$ are the commutants in $E_8$
of $H$ and $H'$, respectively. Let us assume that there are $k_1$ instantons
in $H$ and $k_2$ in $H'$, with $k_1+k_2=24$. The F-theory
dual of this model is provided by a \cyt\ \ca{M}. Then,
$$h_{11}(\ca{M}) =  \hbox{rank}(G) + \hbox{rank}(G') + 3.$$
Applying the map, we find that the proposed mirror model has $k_1$ small
instantons on an {\ss H} orbifold singularity and $k_2$ small instantons on
an {\ss H'} orbifold singularity. $h_{21}(\ca{W'})$ is the number of
deformations of complex structure.
Since all the instantons are now small, and sitting at special points, they do
not contribute any moduli. Fixing the orbifold singularities
in the $K3$ results in a reduction of the number of hypermultiplet moduli
(deformations) by
an amount given by moduli({\ss H}) + moduli({\ss H'}) (where moduli({\ss H})
is the
number of moduli needed to specify the type {\ss H} singularity),
so that we find
$$h_{21}(\ca{W'}) = 20 - \hbox{moduli({\ss H})}
- \hbox{moduli({\ss H'})} - 1.$$
Now, one can verify (\eg by studying the branching rules), that
moduli({\ss H}) + rank($G$) = rank($E_8$) = 8.
Thus we see that
$$h_{21}(\ca{W'}) = h_{11}(\ca{M}).$$

Next, the number of deformations of complex structure of \ca{M} is equal
to the number of hypermultiplet moduli of our original heterotic model
minus one. The latter consists of the moduli of the gauge bundle and those of
the $K3$ itself. Recall that the dimension of the moduli space of $k$ 
instantons in a group $H$ equals $h(H)k-{\rm dim}(H)$, where $h(H)$ denotes
the dual Coxeter number of $H$. Also the $K3$ is generic and hence provides
us with 20 moduli. Thus,
$$h_{21}(\ca{M})=h(H)k_1-{\rm dim}(H)+ h(H')k_2-{\rm dim}(H')+19.$$
On the other hand, 
$$h_{11}(\ca{W')}={\rm rank}(\tilde{G})+\tilde{n}_T+2,$$
where $\tilde{G}$ is the total gauge group and $\tilde{n}_T$ is the total
number of
tensor multiplets resulting from putting $k_1$ point-like instantons on a 
${\ss H}$ singularity and  $k_2$ point-like instantons on a ${\ss H'}$ 
singularity of the $K3$. The outcome of such a process can be easily found
using the results of~\cite{\AM}. We give the results in Table~\tabref{rank}.
From this, it is clear that $k$ small instantons on a {\ss H} orbifold
singularity give rise to a gauge group \ca{G} and $n_T'$ tensor multiplets
such that
$$ {\rm rank}(\ca{G}) + n_T' = h(H)k - {\rm dim}(H).$$ In addition,
we also obtain the primordial $E_8\times E_8$ gauge group (since there are no
large instantons left) and one tensor multiplet, so that
$$h_{11}(\ca{W'}) = h_{21}(\ca{M}).$$

Now consider the situation where there are small instantons sitting
on {\sl smooth\/} points of the $K3$. Smoothing out the singularity due to a
small instanton requires one blowup, so we find that the contribution to
$h_{11}$ is one. Furthermore, a small instanton also yields a single
hypermultiplet modulus, corresponding to its position, hence the contribution
to $h_{21}$ is also one. Therefore, mapping a small instanton on a smooth
point of the $K3$ to itself is consistent with mirror symmetry. Clearly,
the proof above holds even when the heterotic vacuum consists of a mixture
of small and large instantons.

Although the proof above considered only the case of $E_8$ instantons,
the same argument applies for simple Spin(32)$/\IZ_2$ instantons. Once again,
by counting the number of tensor multiplets and gauge groups obtained when
small Spin(32)$/\IZ_2$ instantons coalesce on an {\ss H} orbifold singularity,
it is easy to see that $h_{11}(\ca{W'}) = h_{21}(\ca{M})$ (see
Table~\tabref{rank2}). The case when
more than one small
Spin(32)$/\IZ_2$ instanton sits on a smooth point of the $K3$ is a little
subtle. While there is only one hypermultiplet modulus coming from the
position of the small instantons, we obtain additional neutral hypermultiplets
upon going to the Coulomb branch of the theory. This is because the gauge group
is non-simply-laced ($Sp(k)$ for $k$ coincident small instantons), and the
charged matter content consists of representations of $Sp(k)$ which contain
zero weight vectors, which yield exactly $k-1$ additional moduli when we go to
the Coulomb branch.
\newpage
\midinsert
$$\def\skip{\hskip2.5pt}
\def\t{{\times}}
\def\extraspace{\omit{\vrule height2.5pt}&&&&&&\cr}
\vbox{\offinterlineskip\halign{\strut # height 12pt depth 6pt
&\hfil\skip \eightrm $#$\skip\hfil\vrule
&\hfil\skip \eightrm #\skip\hfil\vrule
&\hfil\skip \eightrm $#$\skip\hfil\vrule 
&\hfil\skip \eightrm $#$\skip\hfil\vrule
&\hfil\skip \eightrm $#$\skip\hfil\vrule
&\hfil\skip \eightrm $#$\skip\hfil\vrule \cr
\noalign{\hrule}
\omit{\vrule height3pt}&&&&&&\cr
\vrule&H&{\ss H}&k&n_T'&\ca{G}&n_T'{+}{\rm rank}(\ca{G})\cr
\omit{\vrule height3pt}&&&&&&\cr
\noalign{\hrule\vskip3pt\hrule}
\omit{\vrule height2pt}&&&&&&\cr
\vrule&SU(3)&{\ss A}$_2$&\ge 5&k&SU(2){\times}SU(3)^{(k-5)}{\times}SU(2)&
3k-8\cr
\extraspace
\vrule&SU(m)&{\ss A}$_{m-1}$&\ge 2m&k&
SU(2){\times}SU(3){\times}\ldots{\times}SU(m{-}1){\times}&
km{-}m^2{+}1\cr
\vrule&&&&&SU(m)^{(k-2m+1)}{\times}\ldots{\times}SU(2)&\cr
\extraspace
\vrule&G_2&{\ss D}$_4$&6&6&SU(2){\times}G_2{\times}SU(2)&4k-14\cr
\extraspace
\vrule&G_2&{\ss D}$_4$&7&2k{-}6&SU(2){\times}G_2^2{\times}SU(2)&4k-14\cr
\extraspace
\vrule&SO(8)&{\ss D}$_4$&\ge 7&2k{-}6&
SU(2){\times}G_2{\times}SO(8)^{(k-7)}&6k-28\cr
\vrule&&&&&{\times}G_2{\times}SU(2)&\cr
\extraspace
\vrule&SO(2m{+}7)&{\ss D}$_{m+4}$&2m{+}6&2k{-}6&
SU(2){\times}G_2{\times}SO(9){\times}SO(3){\times}\ldots{\times}&2m^2+9m+9\cr
\vrule&&&&&SO(2m{-}1){\times}SO(2m{+}7){\times}&\cr
\vrule&&&&&SO(2m{-}1){\times}\ldots{\times}SU(2)&\cr
\extraspace
\vrule&SO(2m{+}8)&{\ss D}$_{m+4}$&\ge 2m{+}7&2k{-}6&
SU(2){\times}G_2{\times}SO(9){\times}SO(3){\times}\ldots{\times}&k(2m+6)-\cr
\vrule&&&&&SO(2m{-}1){\times}SO(2m{+}7){\times}Sp(m){\times}&
2m^2{-}15m{-}28\cr
\vrule&&&&&(SO(2m{+}8){\times}Sp(m))^{(k-2m-7)}{\times}&\cr
\vrule&&&&&\ldots{\times}SO(9){\times}G_2{\times}SU(2)&\cr
\extraspace
\vrule&E_6&{\ss E}$_6$&\ge 10&4k{-}22&SU(2){\times}G_2{\times}F_4{\times}
SU(3){\times}&12k{-}78\cr
\vrule&&&&&(E_6{\times}SU(3))^{(k{-}9)}{\times}F_4{\times}G_2{\times}SU(2)&\cr
\extraspace
\vrule&E_7&{\ss E}$_7$&\ge 10&6k{-}40&SU(2){\times}G_2{\times}F_4{\times}
G_2{\times}SU(2)\t E_7\t &18k{-}133\cr
\vrule&&&&&(SU(2)\t SO(7)\t SU(2)\t E_7)^{(k{-}10)}&\cr
\vrule&&&&&\t SU(2)\t G_2\t F_4\t G_2\t SU(2)&\cr
\extraspace
\vrule&E_8&{\ss E}$_8$&\ge 10&12k{-}96&E_8^{(k-9)}\t F_4^{(k-8)}\t&30k{-}248\cr
\vrule&&&&&G_2^{(2k-16)}\t SU(2)^{(2k-16)}&\cr
\omit{\vrule height2pt}&&&&&&\cr
\noalign{\hrule}
}}
$$
\nobreak\tablecaption{rank}{$k$ small $E_8$ instantons on an {\ss H} orbifold
singularity. Note that each entry in the last column is equal to the
dimension of the moduli space of $k$ instantons in a gauge bundle with
structure group~$H$. Based on Table 2 of Ref.~\cite{\AM}.}
\bigskip
\endinsert
\newpage
\midinsert
$$\def\skip{\hskip2.5pt}
\def\t{{\times}}
\def\extraspace{\omit{\vrule height5pt}&&&&&\cr}
\vbox{\offinterlineskip\halign{\strut # height 12pt depth 6pt
&\hfil\skip \rm #\skip\hfil\vrule
&\hfil\skip \eightrm $#$\skip\hfil\vrule 
&\hfil\skip \eightrm $#$\skip\hfil\vrule
&\hfil\skip \eightrm $#$\skip\hfil\vrule
&\hfil\skip \eightrm $#$\skip\hfil\vrule \cr
\noalign{\hrule}
\omit{\vrule height3pt}&&&&&\cr
\vrule&{\ss H}&k_{{\rm min}}&n_T'&\ca{G}&n_T'{+}{\rm rank}(\ca{G})\cr
\omit{\vrule height3pt}&&&&&\cr
\noalign{\hrule\vskip3pt\hrule}
\omit{\vrule height2pt}&&&&&\cr
\vrule&{\ss A}$_{m-1}$&2m&{m\over 2}&Sp(k)\t SU(2k-8)\t SU(2k-16)\t
\ldots&km - m^2 +1\cr
\vrule&($m$ even)&&&\t SU(2k-4m+8)\t Sp(k-2m)&\cr
\extraspace
\vrule&{\ss A}$_{m-1}$&2m{-}2&{m-1\over 2}&Sp(k)\t SU(2k-8)\t SU(2k-16)\t
\ldots&km - m^2 +1\cr
\vrule&($m$ odd)&&&\t SU(2k-4m+4)&\cr
\extraspace
\vrule&{\ss D}$_{m+4}$&2m{+}8&m{+}4&Sp(k)\t Sp(k-8)\t SO(4k-16)\t&k(2m+6) -\cr
\vrule&($m$ even)&&&Sp(2k-16)\t \ldots \t Sp(2k-4m-8)&
{1\over 2}(2m+8)(2m+7)\cr
\vrule&&&&\t SO(4k-8m-16)\t Sp(k-2m-8)^2&\cr
\extraspace
\vrule&{\ss D}$_{m+4}$&2m{+}6&m{+}3&Sp(k)\t Sp(k-8)\t SO(4k-16)\t \ldots&
k(2m+6) -\cr
\vrule&($m$ odd)&&&\t Sp(2k-4m-4)\t SO(4k-8m-8)&
{1\over 2}(2m+8)(2m+7)\cr
\vrule&&&&\t Sp(2k-4m-12)\t SU(2k-4m-12)&\cr
\extraspace
\vrule&{\ss E}$_6$&8&4&Sp(k)\t SO(4k-16)\t Sp(3k-24)&12k-78\cr
\vrule&&&&\t SU(4k-32)\t SU(2k-16)&\cr
\extraspace
\vrule&{\ss E}$_7$&12&7&Sp(k)\t SO(4k-16)\t Sp(3k-24)\t&18k-133\cr
\vrule&&&&SO(8k-64)\t Sp(2k-20)\t Sp(3k-28)&\cr
\vrule&&&&\t SO(4k-32)\t Sp(k-12)&\cr
\extraspace
\vrule&{\ss E}$_8$&11&8&Sp(k)\t SO(4k-16)\t Sp(3k-24)\t&30k-248\cr
\vrule&&&&SO(8k-64)\t Sp(5k-48)\t SO(12k-112)&\cr
\vrule&&&&\t Sp(3k-32)\t Sp(4k-40)\t SO(4k-32)&\cr
\omit{\vrule height2pt}&&&&&\cr
\noalign{\hrule}
}}
$$
\nobreak\tablecaption{rank2}{$k$ small Spin$(32)/\IZ_2$ instantons on an
{\ss H} orbifold singularity. Note that each entry in the last column is
equal to the dimension of the moduli space of $k$ instantons in a gauge bundle
with structure group $H$. Based on Table 4 of Ref.~\cite{\AM}.}
\bigskip
\endinsert
\newpage
\section{eg}{Examples}
We have shown that the Hodge numbers of the \cyt\ \ca{W'} constructed by
using the large/small instanton map and the heterotic/F-theory
duality transformation are precisely those of the actual mirror \ca{W}. While
constituting a rather strong evidence in favor of our conjecture, this fact
however does not guarantee that \ca{W'} coincides with \ca{W}. To strengthen
our position, we will now make use of toric methods which allow us, given
a \cyt\ realized as a hypersurface in a toric variety, to explicitly construct
its mirror.\Footnote{To a \cy\ hypersurface in a toric variety there 
corresponds a (reflexive) Newton polyhedron $\D$ together with its dual
$\nabla$, which is the Newton polyhedron of the mirror.} Our strategy is 
to take \ca{M} which provides the F-theory dual of a (perturbative) heterotic
vacuum, construct its mirror \ca{W} torically, and compare it with 
\ca{W'} obtained by applying the large/small instanton map. Our task is 
considerably simplified by the fact that the mirrors which we need were 
constructed in~\cite{\BG}, where their massless vector and tensor multiplet
spectra were identified, and the physics of small instantons on orbifold
singularities was worked out in~\cite{\AM}. Thus what we really need to do
is to compare the results of these two references. We list below a few
examples. 

$(1)$ As pointed out in \SS{2}, the \cyt\ with Hodge numbers
$h_{11}=491$ and
$h_{21}=11$ has precisely the singularity structure that is obtained when
24 small $E_8$ instantons coalesce onto an {\ss E}$_8$ singularity of the $K3$.
On the other hand, this manifold is also the mirror of the \cyt\ that yields
the F-theory dual of the heterotic $E_8{\times}E_8$ model with 24 (large)
instantons in a single $E_8$ factor. We can actually go further and study the
actual intersection pattern of the divisors corresponding to the the blowups
of the singularities. This was worked out for the heterotic model
in Ref.~\cite{\AM}. For the \cyt , this pattern can be read off from the dual
polyhedron~\cite{\BG,\Mat}. Once again, we find an exact match for this
example and for all the other examples listed below.

$(2)$ Next, consider the \cyt\ with Hodge numbers
$h_{11}=3$, $h_{21}~=~243$. This
corresponds to the heterotic model with 12 instantons in each $E_8$ factor.
The mirror manifold has Hodge numbers $h_{11}=243$ and $h_{21}=3$, and gives
gauge group $E_8^8{\times}F_4^8{\times}G_2^{16}{\times}SU(2)^{16}$,
and 97 tensor multiplets. Now consider the heterotic model obtained from the
map of \SS{2}. We have two {\ss E}$_8$ singularities with 12 small instantons
on each of them. The total gauge group then consists of the primordial
$E_8{\times}E_8$ in addition to two factors of
$E_8^3{\times}F_4^4{\times}G_2^{8}{\times}SU(2)^{8}$, and $48+48+1=97$ tensor
multiplets, in agreement with the previous result.

$(3)$ Now unhiggs an $SU(2)$ subgroup of $E_8$ in the previous
example. This yields
a \cyt\ with Hodge numbers $h_{11}=4$ and $h_{21}=214$. On the heterotic side,
this corresponds to 12 instantons in an $E_7$ bundle and 12 more in an $E_8$
bundle. From the toric data, we find that the mirror of the \cyt\ gives
81 tensors and total gauge group
$E_8^5{\times}E_7^3{\times}F_4^6{\times}G_2^{12}{\times}SO(7)^2{\times}SU(2)^{16}$. The heterotic model obtained from the map of \SS{2} now gives a total
gauge group consisting of the primordial $E_8{\times}E_8$ in addition to one
factor of $E_8^3{\times}F_4^4{\times}G_2^{8}{\times}SU(2)^{8}$ and another
factor of
$E_7^3{\times}F_4^2{\times}G_2^4{\times}SO(7)^2{\times}SU(2)^8$, as well as
$48+32+1=81$ tensors. 

$(4)$ If we were to unhiggs $SU(3)$ instead of $SU(2)$ in the
previous
example, we would have to put the first 12 instantons in an $E_6$ bundle. The
Hodge numbers of the corresponding \cyt\ are $h_{11}=5$ and $h_{21}=197$.
The toric data reveal
that the mirror of this manifold gives 75 tensors and total gauge group
$E_8^5{\times}E_6^3{\times}F_4^6{\times}G_2^{10}\times SU(3)^4{\times}
SU(2)^{10}$. The heterotic model
obtained from the map of \SS{2} now gives a total
gauge group consisting of the primordial $E_8{\times}E_8$ in addition to one
factor of $E_8^3{\times}F_4^4{\times}G_2^{8}{\times}SU(2)^{8}$ and another
factor of $E_6^3{\times}F_4^2{\times}G_2^2{\times}SU(3)^4{\times}SU(2)^2$, as
well as $48+26+1=75$ tensors.   

$(5)$ Now unhiggs an $E_7$ subgroup of $E_8$ in the example with 12
instantons in each $E_8$. This yields
a \cyt\ with Hodge numbers $h_{11}=10$ and $h_{21}=152$. On the heterotic side,
this corresponds to 12 instantons in an $SU(2)$ bundle and 12 more in an $E_8$
bundle. From the toric data, we find that the mirror of the \cyt\ gives
61 tensors and total gauge group
$E_8^5{\times}F_4^4{\times}G_2^{8}{\times}SU(2)^{17}$. The heterotic model
obtained from the map of \SS{2} now gives a total
gauge group consisting of the primordial $E_8{\times}E_8$ in addition to one
factor of $E_8^3{\times}F_4^4{\times}G_2^{8}{\times}SU(2)^{8}$ and another
factor of $SU(2)^9$, as well as $48+12+1=61$ tensors. 

$(6)$ If we were to unhiggs $E_8$ instead of $E_7$ in the previous
example, we would obtain a model with 12 extra tensor
multiplets (coming from an equal number of small instantons sitting on smooth
points of the $K3$) and $E_8$ gauge symmetry. The Hodge numbers of the
corresponding \cyt\ are $h_{11}=23$ and $h_{21}=143$. The toric data reveal
that the mirror of this manifold gives 61 tensors and total gauge group
$E_8^5{\times}F_4^4{\times}G_2^{8}{\times}SU(2)^{8}$. The heterotic model
obtained from the map of \SS{2} now gives a total
gauge group consisting of the primordial $E_8{\times}E_8$ in addition to one
factor of $E_8^3{\times}F_4^4{\times}G_2^{8}{\times}SU(2)^{8}$, as well as
$48+12+1=61$ tensors, since the 12 small instantons map to themselves, giving
12 tensors and no additional enhancement of gauge symmetry.   

$(7)$ Consider the heterotic model with 8 instantons in an $SO(8)$
gauge bundle and 16 instantons in an $E_8$ gauge bundle. This gives a model
with $SO(8)$ gauge symmetry. The corresponding \cyt\ has Hodge numbers
$h_{11}=7$ and $h_{21}=271$. The toric data reveal that the mirror manifold
gives 107 tensors and gauge group
$E_8^9{\times}F_4^9{\times}G_2^{18}{\times}SU(2)^{18}$. This appears to
disagree with the heterotic result, which gives
$E_8^9{\times}F_4^8{\times}SO(8){\times}G_2^{18}{\times}SU(2)^{18}$ instead.
This is because we need to be more careful in identifying the groups from the
toric data. Out of the nine $F_4$ factors seen in the polyhedron, one is
different from all the rest. The divisor corresponding to this factor has
self-intersection $-4$ rather than $-5$, meaning that there is charged matter
in the {\bf 26} of~$F_4$~\cite{\Mat}. Exact agreement with the heterotic
result can now be obtained by Higgsing the $F_4$~to~$SO(8)$. It is easy to see
that the Hodge numbers remain unchanged during this process. 

$(8)$ Consider the heterotic model with 6 instantons in an $SU(3)$
gauge bundle and 18 instantons in an $E_8$ gauge bundle. This gives a model
with $E_6$ gauge symmetry. The corresponding \cyt\ has Hodge numbers
$h_{11}=9$ and $h_{21}=321$. The toric data reveal that the mirror manifold
gives 127 tensors and gauge group
$E_8^{11}{\times}F_4^{10}\times G_2^{21}{\times}SU(2)^{22}$. Again, this
disagrees with the heterotic result, which is
$E_8^{11}{\times}F_4^{10}{\times}G_2^{20}\times SU(3){\times}SU(2)^{22}$.
Once again, we need to be careful in the analysis of the toric data. One of
the $G_2$'s is seen to be different from all the rest in that there is extra
charged matter in  the {\bf 7} of $G_2$~\cite{\Mat}, which can be used to
Higgs the $G_2$ down to $SU(3)$, to obtain exact agreement with the heterotic
result.

$(9)$ Finally, consider the self-mirror manifold with Hodge numbers
$(43, 43)$. The corresponding heterotic model may be obtained in two ways.
One is to consider the $E_8{\times}E_8$ model with 24 tensor multiplets, and
the other is to consider the $SO(32)$ model with 24 coincident small
instantons which gives an additional $Sp(24)$ gauge symmetry. It was argued in
Ref.~\cite{\AM} that these models are T-duals of each other. In each case,
the heterotic model maps to itself under the map of \SS{2}, in agreement with
the fact that the manifold is self-mirror.
\newpage
\section{fin}{Discussion}
In this note, we have proposed a map relating mirror symmetry in \cyt s to
heterotic compactifications. We have shown that the heterotic equivalent of
the mirror map relates large instantons in a gauge bundle
and small instantons on the corresponding orbifold singularity. We find that
this map is consistent with the exchange of Hodge numbers under the mirror map,
and have also shown that it is consistent with the singularity structure of
the manifolds in a number of examples. This is strong evidence in favor of the
proposal.

Although we have mainly considered the $E_8\times E_8$ heterotic
theory, our results are applicable to the $SO(32)$ theory also. While
we have proved that the Hodge numbers match irrespective of whether the
instantons are of $E_8$ or Spin$(32)$ type, comparing the singularity structure
is a different matter. The singularity structure of the mirrors of the \cyt s
corresponding to $SO(32)$ models are currently being worked out~
\Ref\CH{P. Candelas and H. Skarke, to appear.} . Preliminary results appear to
corroborate our statements, though more examples need to be worked out. Of
course, if we were to compactify further on a circle, the resulting T-duality
between the $E_8\times E_8$ and $SO(32)$ theories would suggest that our
results should also hold for the $SO(32)$ models.
 
These results should hopefully shed some light on the origins of mirror
symmetry in \cym s. While our results hold
only for elliptic \cyt s whose mirrors are also elliptic fibrations (because
F-Theory/Heterotic duality holds only for elliptic fibrations), perhaps a
generalization of this approach works for all \cym s\Footnote{It would be
interesting to see how our results are related to the interesting proposal of~
\Ref\SYZ{A. Strominger, S.-T. Yau and E. Zaslow, \npb{479} (1996) 243,\\
hep-th/9606040.}\ that mirror symmetry is in fact a form of T-duality.}.

\vskip5pt
\noindent {\bf Acknowledgements}
\vskip5pt
\noindent We wish to thank P.~Candelas for reading a preliminary version of
this paper and providing useful comments.
This work was supported by the Robert Welch Foundation and NSF grant
PHY-9511632.

\newpage
\immediate\closeout\referencewrite\referenceopenfalse
\line{\bf\hfil References\hfil}\bigskip\parindent=0pt\input referenc.texauxil
\bye